# Developing industry-wide information management capabilities: A case study from British Columbia's tree fruit industry

**Svan Lembke, Youry Khmelevsky and Lee Cartier**

**Okanagan College, British Columbia, Canada**

## Abstract

As an industry of primarily small and mid-size businesses, it is becoming increasingly more difficult for British Columbia (BC)'s tree fruit growers to compete with large, often vertically integrated producers from other regions. New ways of managing information and resources collaboratively are needed to develop competitive strengths. This case study seeks to understand the information and knowledge management capabilities of the BC tree fruit cluster across the value chain for six different data domains. A qualitative methodology design of 21 in-depth interviews with cluster stakeholders provides insights into the data quality, completeness and integration points, and then applies CMMI level criteria to assess the information management capabilities of the industry. Significant data and process gaps are identified. This paper explores how the BC tree fruit industry can move forward from this position using technology solutions to support the development of information and knowledge, and collective decision-making.

# 1 Introduction

## 1.1 The importance of information management in agriculture

Recent publications have tried to review and make sense of the research into agricultural systems and innovation [1]–[3]. Sophisticated technology and complex innovation approaches are the focus of many research projects [4]–[9]. The degree of technological sophistication poses a challenge when an agricultural community such as the British Columbia (BC) tree fruit industry seeks guidance with their attempts to innovate.  It is unclear where the agricultural community should start. This paper does not wish to replicate the excellent and in-depth review of existing research into innovation systems, or add sophisticated technology solutions for agriculture but focuses only on the creation of an enabling digital platform for collaboration and innovation in one small industry in BC, Canada.

There is an increasing acceptance that most innovation in today's global economy is intimately connected to technology and data[8][10]. Information technology has become a services science that facilitates innovation across a multitude of industries [11].  Research into knowledge management shows that how a company uses its knowledge often becomes its competitive advantage[12]. Agricultural industries are said to benefit significantly from the use of data [4]. Businesses in agriculture have been given many tools but adoption and innovation often seem to lag. The investment into information technology services to assist with tracking, analysis, and decision-making, is slow even in Organization for Economic Collaboration and Development (OECD) countries partly due to the high upfront cost and slow return on investment [13]. Furthermore, information delivered through information technology alone is insufficient to drive innovation, it still requires strategic decision-making.  According to [13], there is a focus on agriculture and the industry-wide adaptation to climate change and innovation in countries like Italy, France, Switzerland, and the Netherlands. Research in [13] uncovered how innovation in today's agriculture is often quite radical and requires breaking with existing production practices.  This makes it difficult for industry stakeholders that seek a gradually adoption path as sought in agricultural innovation system theories referred to by in [1].

There is a need for inclusive processes and more understanding of how to organize the systems of change and innovation[14], especially as we strive for sustainable agricultural practices[15]. Digitalization and use of data could be a transformative force but there is limited research that can prove this[3].  In agricultural sectors where the value chains are often split among a multitude of small and mid-size organizations, having information in digital form seems essential for its exchange and analysis.  This can either be in big corporations or information systems that connect a multitude of stakeholders [8].  In [16],  the process of up-scaling (change within an institution) and out-scaling (change across institutions in agricultural networks) and their respective innovation methodologies are described, starting with niche practices and gradually translating them to a broader environment. Connecting a multitude of stakeholders is likely to require both.

In British Columbia (BC) the tree fruit industry with 15000 acres and around 400 small and mid-size commercial growers [17] is suffering from a multitude of challenges, making the growing and selling of tree fruits, primarily apples and cherries, increasingly difficult and affecting the ability to innovate and maintain profit margins [18].  Information and information management processes appear disjointed despite the logic that this could deliver impressive gains in competitiveness as a direct function of the knowledge and decision-making capabilities created along the value chain. This is consistent with findings by [1] on agricultural state-of-the-art systems.  Information management within organizations and across organizations is needed for transparency of production methods across the value chain.  This paper seeks to translate some of the research findings into a roadmap for integrated information management using the BC tree fruit industry as a case study.  Findings from this research may provide insights for small agricultural industries in other regions.

### 1.2    From information architecture to innovation network

Information technology is a fast-changing field of science. Information technology seeks to create systems that prevent redundancies, ensure data validity and integrity over time and across different entry/interface points, and reliable methods of analysis and reporting [11]. There are multiple roles for information entry, validation, approval and use within an organization or industry, and this introduces differences in authorization and access.  Proprietary data protection and entry of aggregated and detailed data must be enabled to create trust and encourage collaboration across stakeholders [20]. This means a 'robust' architecture is needed that can ensure data quality over time (avoiding redundancies) and cope with changes in data entry methods and changing relationships between data tables that result from changing production practices, including disruptive practices. For the BC tree fruit industry, this will inevitably include visibility of data of fruit production for specific markets and their requirements.

Previous research by [19] into the BC grower community and their technology readiness confirms that much of the data collected by growers is not digitized. There are many intermediaries that contribute to grower decision making but their effectiveness is limited as a direct result of the data available.  In follow up to these findings, this study intends to take a step back from aspirational precision agriculture innovation and big data opportunities, and seek to understand what kind of data and systems are in place in the industry, where and how they can be improved and transformed to better assist the grower community. The primary focus of this project is to determine the data requirements and the logical connections of the information across the industry, independent of the choice of present or future physical databases, software systems or agricultural innovation plans.

The data and data integration points are referred to as an 'architecture', whereby a structure is imposed to understand the building blocks and contact points to collect, connect, analyze and share valuable information. The BC tree fruit industry has large amounts of production data, as well as models for integrated pest management purposes [19]. A system of industry information and innovation requires effective analysis and integration across different data sources and knowledge chains.  The knowledge pyramid found in [21] highlights the need for raw data as the foundation for information and knowledge processing in any information architecture. In agricultural industries where there is a multitude of stakeholders, the concept of knowledge management and innovation requires a cluster-wide systematic effort. The challenge is to build a system that supports this effort. Such a system requires superseding levels of well-defined integrated processes and technology [21], such as defined in the capability maturity model integrated (CMMI) which traditionally deals with the way an organization must follow to establish and maintain well-mapped processes[12].  This model is expanded here to apply to the integrated processes of an entire industry.  The CMMI has succeeding stages of evolution such as found when an organisation (or industry) matures and is able to consistently repeat a process and achieve high quality. The CMMI approach thus provides an assessment tool to guide the process from the industry merely collecting data to the wisdom generated from it as per the knowledge pyramid.

| CMMI Level | Maturity Description |
|---|---|
| Level 1 | This is the **initial level**. The activities and processes at this level have significant gaps, are unpredictable, poorly controlled, and reactive. |
| Level 2 | This level shows some degree of management. The processes are still specific towards certain activities or projects and are often reactive. This level starts to define the basic management of data at a **consistent and repeatable** level. Most industries to date do not have integrated process that achieve a Level 2 or higher. |
| Level 3 | This level assumes clearly defined comprehensive processes. Processes are **proactive, formalised and standardized**. This is the level that assumes availability of information that can be used as a foundation for knowledge management, as illustrated by [21]. Large dominant corporations can facilitate a Level 3 for the industry but this level is difficult to achieve for any industry of small and mid-size organizations. Once established an innovation network is in place. |
| Level 4 | This level has quantitatively managed processes. The broad range of processes are measured and controlled, with high degrees of quality assurance. Knowledge management is thus available to the users and defines a degree of **best practise**. Best practise infers decision making which confirms the effectiveness of the innovation network. |
| Level 5 | This level provides the organization or industry with the ability to review their processes and continuously **optimise them and innovate**. Knowledge management has achieved a level of wisdom and thus the ability to develop new connections to the benefit of the network. |

*Table 1: CMMI Level descriptions for industry assessment adapted from Dayan & Evans (2006)*

The CMMI differentiates between five maturity levels. Table 1 describes the different level with reference to organizations and the industry. The transition from Level 1, the entry level, to Level 2 requires the digitalization of data and seeks to established niche pockets of quality data. Over time this can become a pool of historical data for analysis and reliable use. At Level 3 best practises for a wide range of data collection processes are agreed and implemented.  Level 3 should be aspired to by agricultural industries as this constitutes an industry information architecture that is ready to support an innovation network across the cluster. This research seeks to understand the scale of effort required to establish such an information architecture for the BC tree fruit industry.

## 1.3     Research questions that guide the mapping of the BC Tree Fruit information architecture

This case study describes the process and findings when mapping an information architecture for the BC tree fruit industry. The importance of an integrated information architecture has become clear from the literature. The digital robustness and scalability of such an architecture is at the forefront of this study. The first step in this process is the assessment of the information to date and development of a possible progression path of the information architecture over time. The assessment process is guided by the knowledge pyramid and CMMI framework, then using this information to draw out gaps and assumptions that determine how the information architecture and systems can be improved.

The research questions (RQ) of this study are:

*RQ1. What information is presently collected and shared by industry stakeholders?*

*RQ2. How can the collection and sharing of available information be optimized within one integrated industry architecture of data storage and supporting processes?*

## 2 Methodology

The research design for this study consists of in-depth structured interviews with 21 participants from the BC tree fruit industry cluster. Based on Porter's (1998) diamond framework, the cluster is defined by the strategic alignment of four determinants, being the demand conditions, the rivalry or strategy between the rivals (the 'growers'), the supporting industries (packing-houses, field services, etc.), and the factor conditions (industry associations and government support groups). As this study follows on from a previous project into the technology readiness of the BC grower community [19], only three progressive growers were included in this research project to represent the cluster. The focus of the investigation is on the supporting industries and factor conditions. This is in line with another[1] review of agricultural literature, confirming the importance of agricultural advisors and intermediaries to the innovation capabilities of the agricultural sector.

The structure of the questionnaire and list of participants was determined during a workshop with five engaged industry stakeholders in June 2019. During the workshop the cluster relationships, the main data requirements and data flows along the value chain of the tree fruit industry were confirmed. Analysis of the workshop data identified that data collection and data management processes can be grouped into six main data domains. A questionnaire was designed for structured face-to-face interviews. The interview questions are organized into the following sections: organizational data; raw data capture, storage and sharing; predictions and decision-making; knowledge development, diffusion and expectations for the future.

The participants in this study represented their organization, or a specific department of their organization. Organizations ranged from growers of apples, cherries or organic tree fruits of different sizes and vertical integration, also packing-houses, suppliers of agricultural goods and services, provincial and federal government departments, research and education institutions and trade associations. They participated in the workshop and/or in a 90-180 minute long interview with at least one of the researchers. A total of twenty-one interviews took place from November 2019 until February 2020. The BC tree fruit industry has approximately 40 organizations supporting the grower community. This study involved more than 40% the supporting organizations.

Extensive qualitative data was collected. All interviews were recorded and transcribed as recommended by [28]. This enabled the systematic coding and assessment of answers for structured analysis. The main focus was on identifying and describing the data collected and used across the industry, also assessing them according to CMMI levels.

# 3 Results and Discussion

The purpose of this research is to map out the information architecture for the BC tree fruit industry. The research findings are discussed in two sections, each focusing on one of the two research questions, starting with the description of the data and information available, then providing a description of the stakeholder landscape and processes for improved integration of information flows.

## 3.1 What information is presently collected and shared by industry stakeholders?

This section outlines the different types of data that is collected and shared across the BC tree fruit industry. The focus is on the type of data, also the format (digital or analogue), consistent quality and use for analysis, also if it is shared and for which purpose.

No single comprehensive repository of data across the industry is available for any data domain. There are however many small data repositories with some analysis. Most participants to this study acknowledged that there is little consistent quality data for analysis at the grower level or for sharing across the industry. The lack of government extension services to assist with data capture, analysis of data and sharing with growers was a frequent discussion point during the interviews. The lack of extension was not further explored as the purpose of this research is not to investigate why or how extension services are provided, but merely to assess and map the data, information architecture and opportunities for the BC tree fruit industry.

### 3.1.1 Geography, Land-use & Plantings (Trees)

The data captured in this data domain includes government census data and a land use survey every five years. The land use survey provides tree fruit commodity and planting density information. Whilst not publicly available, it feeds models such as a water demand calculator that are available online. If the survey results were available publically, it is unclear how they would be used to assist growers. The survey results are primarily for the research community and government departments. Additional research is commissioned to diagnose and predict land-use changes. The main purpose is to look at trends, not detailed grower behavior nor to identify best practices.

Growers capture raw data about their own orchards (at varying degree of detail and quality). Data collection is sometimes carried out by support organizations or consultants that provide soil tests or irrigation designs. Planting and production data are also captured but not in digital format and not analyzed over time. For growers that ship fruit to a packing-house, some high level data of their plantings is captured and shared but it is usually not in a format that makes regular analysis of the data possible.

Other smaller initiatives such as tree canopy assessment, re-plant applications, ad hoc grower member surveys and forms for crop insurance are taking place. None of these are integrated or analyzed for industry use. Sharing of this kind of data is possible upon request but rarely requested by anyone. The only main data gap identified by participants is the absence of a database with land-use and grower contact details. It is difficulty to keeping contact details of industry stakeholders up to date, and thus challenging to communicate with the right people especially in times of emergency. At present this is the responsibility of each individual organization with much room for error and duplication.

In summary whilst there is some good quality industry data in digital format, the digitized data is at high level. There is limited detail about the tree age and replanting information, specific varieties and density to capture tree fruit specific industry details. The lack of detail and consistency across the industry for this data domain can only be described as CMMI level 1. The land use survey every five years can be considered CMMI level 2 but at this stage, it is only of very limited use to the grower community.

### 3.1.2 Water Use and Irrigation

Water use and irrigation is a data domain of much debate as there are concerns about the availability of regional water for agriculture. At the time of this study, agricultural water use is metered but not in real time, not online and not with detailed up-to-date reports. The charges for water have not increased significantly over the past years. Irrigation systems have improved, and industry-wide water consumption has not seen much increase, thus few people request better recording and management of water. Some participants believe that this is likely to change over the next few years if water pricing changes. In the meantime, growers manage their own irrigation systems at varying levels of sophistication and very limited data is collected.

In summary, there is aggregated historical meter data of water consumption and there are some sophisticated models for calculating water use developed by government organizations. There is limited need or use of the available water demand models by growers and there was skepticism about the accuracy or usefulness of these models. In light of the grower community, there is no detailed data on water consumption for different parts of their orchards and thus no historical data for analysis to guide decision making about future water use. Until water becomes a more precious commodity, the lack of detailed data is unlikely to change. The CMMI level of this data domain across the industry can only be described as CMMI level 1.

### 3.1.3 Weather

Timely and accurate weather data is essential for most agricultural industries.  The market for weather services is growing as inter-disciplinary, site-specific information has become critical for effective decision making [26]. Few BC tree fruit growers operate at a scale where they can afford their own weather station(s).  However, tree fruit growers have access to regional weather stations through a range of government-owned and government-funded organizations. The largest network of weather stations, placed on Okanagan agricultural lands is managed by a for-profit grower supplies retailer. Access to their weather information is free for grower members of the biggest cooperative packing-house in the region, and a chargeable service for others. It is also an integrated part of a free online tree fruit pest and disease management system, the Decision Aid System (DAS).  Weather data quality is generally praised but there are concerns about the long-term interest of the retailer to manage this network.

Weather data becomes valuable when it is used for diagnosis and predictions. The predictions for tree fruits because of weather are carried out by individual growers, their advisors or are taken from the DAS to determine the need for spray activities.  Diagnostics executed by government scientists tend to focus on longitudinal and climate change data reported in scientific studies and documents. There is limited focus by growers on long-term weather patterns, except for when deciding on expanding cherry plantings to previously unsuitable land.

The weather data domain is the area receiving the most recommendations for improvement. The main concerns are the accuracy of data and the reach of the network, but not the use of the data for meaningful analysis and decision-making. Whilst the data collection process is consistent and repeatable, the management of this data by growers is not. The CMMI level of this data domain across the industry (not the management of the weather stations and data by the retailer) is therefore described as between CMMI level 1 & 2. The long-term viability of the weather network is not taken into consideration in this assessment.

### 3.1.4 Fertilization, Nutrients, Pest & Disease Management

This is the data domain where growers collect most of the raw data and thus a large amount of data should be available. Fertilization, nutrient, pest and disease management is one of the most complicated areas of tree fruit production and detailed quality data is important. The capture and reporting of some of this data is a requirement for selling the fruit. There are several processes that contribute to the collection of data:

- Health and safety forms capture data about the chemicals growers used. These are required by the packing-houses and retailers. Most growers submit them in hard copy or as pictures of hard copies. These are stored and not analyzed. No funding is available to convert the data into different digital formats and there appears to be no analysis.
- There should be other data that is recorded by growers but this is usually in hand-written format and few growers have the time or resources to enter it into a digital system for subsequent analysis. The quality and detail of this data varies across the grower community.
- Growers that export cherries collect detailed pest trap data as well as maintain spray records. Manually recorded information is transcribed into spreadsheets by the trade association for the Canadian Food Inspection Agency (CFIA). The trade association also collects export specific information for certification of its members.
- Other raw data, such as the purchase of pesticides and fertilizers, is captured by the retailers as these are often regulated items. This information is stored as part of the purchase records for growers.
- The BC government and regional taxpayers fund a sterile insect research (SIR) operation that collects 20 weeks of information about one specific pest (Coddling Moth) for apple and pear orchards. There is one trap per hectare. A new database and online updates of trap results have recently been made available to the public and growers.

The important part of this data domain is its predictive ability. There is a large amount of research taking place by government organizations to produce information about new pests, diseases and spray recommendations for a wide variety of tree fruits. The purpose is to identify pests early and intervene before it causes significant harm to the harvest. The data used for research in this area is project-based and does not rely on industry-wide current data. The research findings and models are consolidated annually to create a document called the 'Production Guide' with spray schedules for growers in subsequent years. This guide is available in hard copy or PDF, and fed into the DAS. The DAS is a system developed by the University of Washington [22] which takes weather station data to assess the upcoming degree days. It has been adjusted for BC and uses the information from the Production Guide to make recommendations about spray requirements for different crops. These recommendations are available to all BC growers at no charge. There is also another federal government-funded advice service. This online system is again free for use but not as detailed and is not specific for tree fruits in BC.

Please note that other scientific research is carried out by government research operations to assist with fertilization and nutrients. The raw data is project specific and not shared, but the analysis and models are shared at meetings, on trade association or government websites, and newsletters.

The DAS, although limited in its scope, is praised by many stakeholder groups. The main challenge is seen in its adoption rate by growers. Also, the future increase in pest threats and market health regulations are considered a challenge by many. Industry-wide actions to adapt and adhere to these changes are considered important for the future success of the industry as a whole. With only 30% of growers using the DAS and only 17% keeping a digital record of their spray activities [19], this data domain can again only be described as CMMI level 1. With a higher adoption of available pest management tools and integrated reporting of spray actions, this could become CMMI level 2.

### 3.1.5 Fruit Production, Sales & Marketing

Detailed fruit production data is collected by growers and at the packing-houses. The growers collect raw data when they estimate their production for the season and then capture the actual harvest quantities and quality (grading). The data is rarely digitized and is at varying degree of detail and accuracy. Fruit production data is usually considered confidential but for those growers that ship to a packing-house, high level trending is sometimes done by the packing-houses. Most of this data is aggregated by the packing-house and reports are provided to growers. These reports are not in digitized format that would enable growers to carry out their own acre specific analysis and decision-making.

There is anecdotal information about new weather and production patterns. Understanding these new patterns is important to stay competitive in future years. Integration of data along the value chain is necessary for interpretation and decision making about crop management practices. However, the existing aggregated nature of raw data, different formats of the data and low quality of the data cannot be used for this purpose.  A CMMI level 1 is achieved for this data domain.

A major gap in data is reported about the marketing of fruit. The marking of fruit and market intelligence could have featured as a separate data domain if more data and processes were available. Trade associations purchase market analysis reports and share findings at member meetings, online or through newsletters. However, these reports only capture snapshots at one particular time and location, with no follow up to analyze the effectiveness of marketing activities or improve these over time. There is a noticeable absence of marketing data for exported apples. With the recent increase in production of Ambrosia apples, this will be critical to enter new markets. The lack of established marketing practices for systematic data-based decision making, and analysis appears to prevent industry-wide discussions and new export programs.

Whilst the lack of market intelligence is raised frequently during the interviews, few anticipate a change in practices despite its apparent need. The industry is described as fragmented and this is influencing the appetite for collective investment in new market entry projects especially for apples. Apples are different to cherries and many other fruit, as retailers in most markets expect growers to supply apples throughout the year.  This requires advanced controlled atmosphere (CA) storage. These storage facilities are expensive to set up and maintain.  It is not feasible for small individual growers to have these facilities. It is important for growers to pool resources and sell collectively.

### 3.1.6   Business and Other

Growers manage their orchards as a business in varying degrees of detail and sophistication, ranging from hand-written notes to expensive integrated business solutions. This is in addition to what they are legally required to report for insurance and tax purposes. This business information is considered highly confidential and will never be shared across the industry.  However, there is a general belief that there is a lack of returns on the fruit, and that the cost associated with some government-mandated production practices are not imposed on the competition that imports fruit into BC.  No data or production analysis was available to show the impact of specific farm management practices on production.  Little interest is expressed in learning and thus collecting data about the business beyond data about labour hours for payroll and other expenses.

Researchers in the past have collected business data from selected growers for analysis of the economic value chains and business optimization projects. The reports of their findings, available to the industry, have had no follow-up by the industry. There is a major gap in understanding growers' cost of production. This prevents industry experts from giving advice to growers about improvements for efficiency and prioritization of investment.

Scientific research projects for new variety breeding programs or new storage, packaging or transportation methods are being carried out. The data is often experimental. Again, findings and analysis are shared at meetings or via reports. There are no extension services but there are some government-subsidized programs to encourage re-planting of trees or adoption of new technology.

Participants to this study believe the industry suffers from a lack of long-term planning about new fruit varieties or production methods. Whilst government funding is available, a lack of understanding and confidence in the proposed actions is preventing industry decision making and adoption. The CMMI level of this data domain across the industry is defined as CMMI level 1.

### 3.1.7   CMMI Assessment of the Data Domains

In summary, there is much data collected across the industry.  It becomes apparent that while government departments collect data and have good databases, their data does reach grower level detail nor is it integrated across the value chain. The format of the data that is available at the grower and packing-house is of varying degree of detail and accuracy, also often not digitized and thus cannot be integrated and used for industry analysis.  Table 2 below summarizes the different CMMI levels across the data domains. The biggest data gaps of the BC tree fruit industry are in the domain for 'Fertilization, Nutrients, Pest & Disease Management' and 'Fruit Production, Sales & Marketing'.

| Industry Data Domain | CMMI Assessment |
|---|---|
| Geography, Land-use & Plantings (Trees) | CMMI level 1 & 2 |
| Water Use & Irrigation | CMMI level 1 (significant data gaps) |
| Weather | CMMI level 1 & 2 |
| Fertilization, Nutrients, Pest & Disease Management | CMMI level 1 (& partially 2) |
| Fruit Production, Sales & Marketing | CMMI level 1 (significant data gaps) |
| Business & Other | CMMI level 1 (significant data gaps & highly confidential) |

*Table 2: Industry data domains & CMMI level assessment*

## 3.2 How can the collection and sharing of available information be optimised within one integrated industry architecture of data storage and supporting processes?

This section strives to understand how the data across the industry is linked for analysis and sharing. This requires the identification of the different data-sources and user groups, and assessment of possible connection between them.

### 3.2.1 Collection of data by stakeholder groups

Having assessed the different data domains across the industry, the data is now analysed by stakeholder group. A fragmentation of data sources is apparent. The data and analysis tools held by the different stakeholder groups can only be referred to as data 'islands', not data domains. The reference to 'islands' recognises the incomplete nature of the data and differences in quality as previously described. Figure 1 illustrates the different data islands, mostly without aggregation nor integration across stakeholder groups. This highlights the following shortcomings:

- The grower community collects the largest amount of data but primarily for their own individual orchards. As previously described, this stakeholder group has data that is inconsistent, incomplete and cannot be aggregated in its present form.

- Government organizations have great breadth of data and there is duplication with the data collected by growers. Comparison or consolidation of the data across stakeholder groups is impossible at this stage due to the different forms of data, detail and categorization.
- Selected service providers hold a multitude of data islands, often on behalf of individual growers. No digital integration points are available across the value chain.
- Scientific researchers collect their own data for diagnosis and provide intermittent advice or models to growers. The models are not validated with actual grower data across the industry.
- Water providers have limited real time data about water consumption and it is presently unconnected to the operational processes of the grower community.
- There is no digital industry platform for communication across stakeholder groups but there are websites and industry meetings to share information.

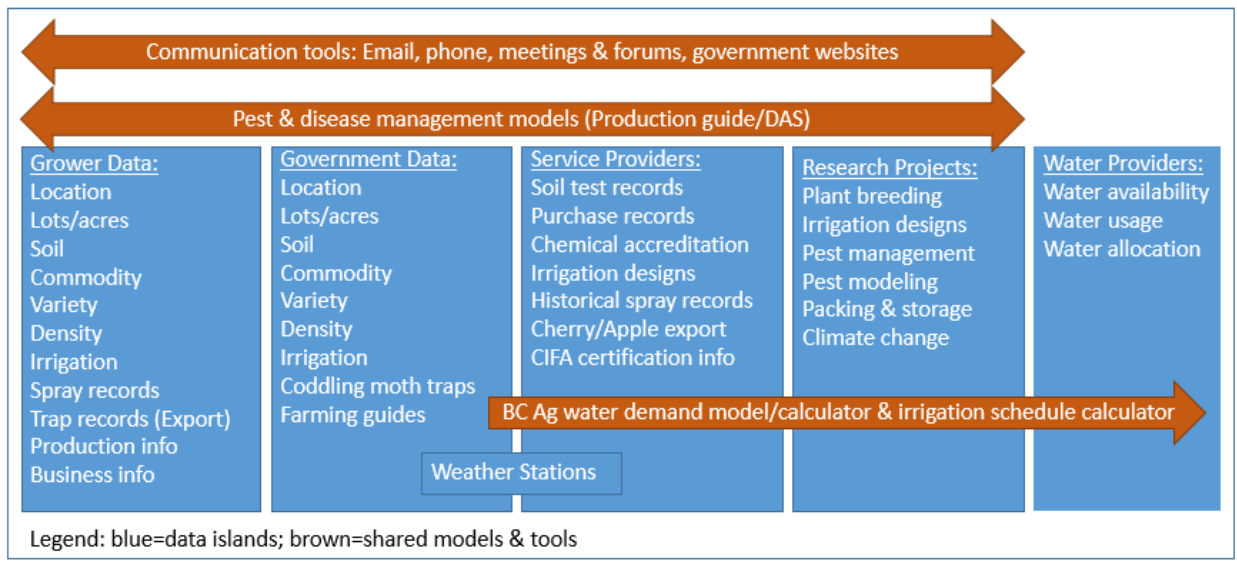

*Figure 1: BC Tree Fruit Industry Data Islands in 2020*

### 3.2.2 Assumptions about an integrated Industry information architecture

The industry tools and models that convert industry data islands into 'information' similar to the second level of the knowledge pyramid [21], lack a strong foundational level of data. The pest and disease management models and the water demand and irrigation calculator (both illustrated in brown in Figure 1) are available for use but the lack of grower data prevents the higher level information processing required in the knowledge pyramid. The following opportunities for improvement can be assumed:

- Agree and prioritize industry best practices for data capture processes at grower level across the industry. This will facilitate the creation of historical data of similar quality and detail for aggregation and analysis. Systems and applications are needed to facilitate easy data entry for growers whilst working in the orchard without requiring re-entry in an office environment. The analysis can enable gradual improvements in grower practices and ease of adaptation to unexpected climate and pest events.
- Integrate packing-house data with grower data where possible to enable analysis across the value chain. This will be critical to connect pack-out reports and sales information to specific acres of production.
- Align grower data with government data collection programs. In addition to improving the quality of data, this could potentially remove significant duplication.
- Make available models (eg Production guide and DAS) an integrated part of the decision-making and data capture process of growers. This will improve grower practices and adherence to guidelines, production reporting and long-term analysis and improvements.
- Invest in market intelligence for established and new markets and collect on-going marketing and sales data for analysis and improvements. These could then also be connected to production data from growers to confirm market requirements are met.
- Manage contact details and communication processes across industry stakeholder groups. This could improve the diffusion of knowledge and the response rate to unexpected climate or pest events.

Prioritizing best practises is challenging for many organizations with hierarchical decision-making processes, let alone a largely unregulated industry. The use of a commercial software solution may guide the process and detail of data entry. A pilot test with a small number of growers using 'Crop Tracker' was underway during the time of this study. Crop Tracker is a commercially available farm management system [24]. Some of the Crop Tracker data entry screens and processes for spray records and production records could be useful starting points to agree data quality and detail, also have easy entry screens on mobile devices. A system like Crop Tracker can instill some of the practices and disciplines required for CMMI Level 2 data capture and processing.

A commercial farm management solution is not an integrated industry information architecture but merely one component that enables individual farm management, not industry management. The following technical considerations appear critical before proceeding with a commercial farm management solution to collect grower data and use it across the industry:

- Grower data must be private but integrated across the value chain with suppliers and packing-houses who are all using the same commercial solution.
- Where approved, data has to be aggregated and available to stakeholder groups outside the value chain and outside the commercial solution.
- Ownership of data and management processes for the entire industry must be decided and executed outside the commercial solution.
- The models and systems available to the BC tree fruit industry (eg DAS) must be integrated with the grower processes.
- Commercial solutions are likely to offer significantly more functionality than will be needed by the BC tree fruit industry (especially at the onset). The cost of gaining access to an easy data collection and repository of data system by using a comprehensive commercial solution must be understood and carefully assessed. This is beyond the scope of this research project.

### 3.2.3 A four-layered technical architecture design

This research began with the intention to map and propose a technical design for an information architecture for the industry. This goes beyond data collection via a commercial solution used by a multitude of growers and packing-houses, but enables industry-wide analysis of data towards industry decision making and allows intermediaries from other stakeholder groups to be involved.

From a technical architecture perspective, a four-layered architecture based on the design principles seen in figure 2 describes the different evolutionary stages of this architecture. The reason for adopting a four-layered design is that it matches the traditional communication and structure of organizations. The design scope of this study applies these four layers to the data and knowledge of the industry, not one organization. The layered design describes simplified functions of the architecture without reference to the content of the data domains. It separates the Presentation Layer that delivers the information to users, from the Business Layer with the logic of the actual business systems within the industry. It then offers a Persistence Layer for pulling and pushing data across systems, and all is anchored in a Data Layer for storage of data within and across the data domains and systems of the industry.

The present information systems environment as informed by the participants to this study cannot be considered a technical architecture as all the layers are incomplete. The end users, primarily growers, are using a multitude of independent and unconnected interfaces, mainly hand-written notes, printed forms and spreadsheets, some get advice from the DAS directly but without a corresponding record of data. Without a digital data foundation, the Persistence Layer cannot engage and share information across systems or across stakeholders.

Optimization of the industry information architecture requires an improved raw data collection and aggregation process. One of the main transformations could be the adoption of a commercial farm management system such as Crop Tracker by growers and packing-houses. Crop Tracker is in modular format that can be integrated with other systems subject to vendor agreement [24][25]. It should be emphasized that Crop Tracker is not the industry platform. Crop Tracker only provides the process and tools for stakeholder data collection and individual business level reporting. The industry still needs its own platform. This can be as simple as building a data warehouse where data from grower systems such as Crop Tracker and other systems is collated and integrated [23].

Such an industry data warehouse must be managed by a knowledgeable and trusted organisation within the cluster, likely a non-commercial entity from the supporting industries. Their role would be the management of the raw data from stakeholder systems, maintenance of the warehouse and also simple industry-level reports. This is captured within the Data Layer of the architecture design. The Persistence Layer actively pulls data into the industry data warehouse and other systems. Future industry requirements may lead to the adoption of more sophisticated business intelligence (BI) applications for the industry, positioned in the Business Layer. The benefits of the industry data warehouse and subsequent industry reporting are control over the data, the ability to integrate and share data across the industry, and analyse and report at industry level. The transparency of production methods across the value chain can be analyzed to deliver opportunities for cost savings, skills sharing and unifying resources to effectively produce for existing and new markets, and much more. Compiling meaningful historical data for industry analysis of this nature is likely to take at least five years.

Figure 2 shows the suggested design of such an information architecture for the industry based on several years of historical data and a digitally integrated systems approach. This will enable researchers, technicians and data specialists to engage with the grower community and each other to provide advice and guidance. Decision-making processes across the industry value chain are supported by the richness of data, multiple data flows, and analysis provided by the integrated systems functionalities. This is unlikely to be accomplished in less than ten years, hence planning for this industry transformation is time sensitive.

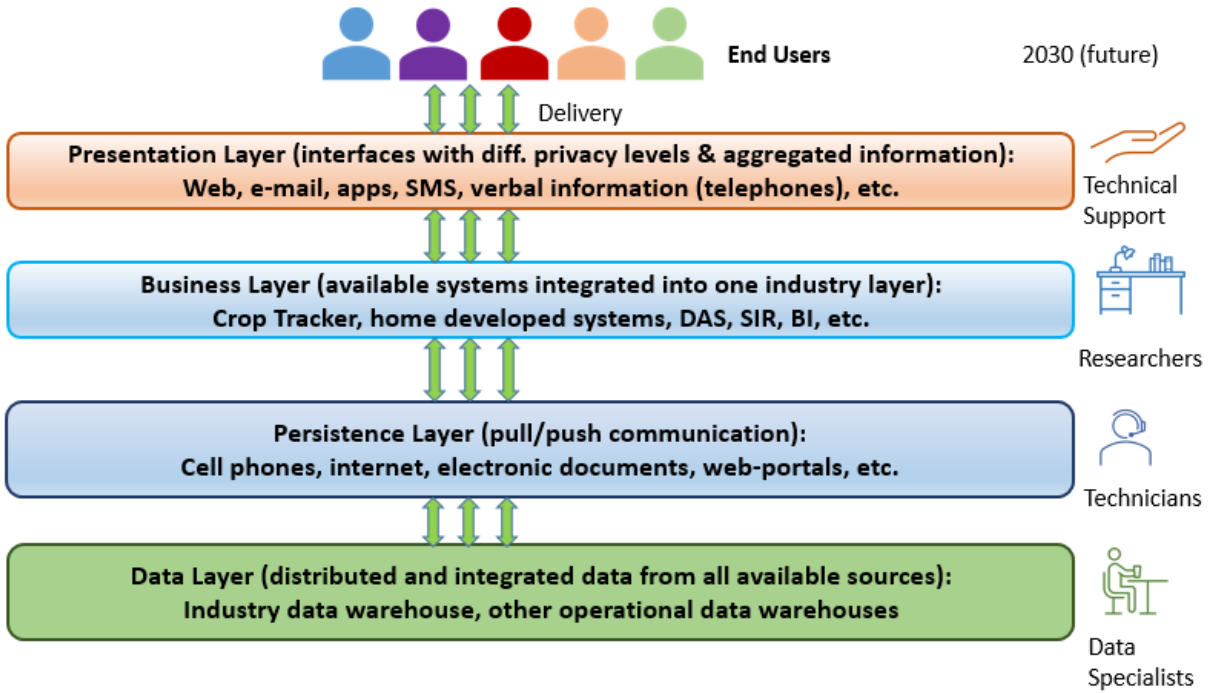

Figure 2: BC Tree Fruit Industry Information Technology Architecture in 2030, adapted from Richards (2015)

Critical design considerations for such a layered design are its scalability and robustness, and the ability to have different privacy settings. The design shows a Data Layer with a data warehouse that stores, connects and integrates data from multiple stakeholders across the value chain, primarily production, packing and market data but potentially also weather and water data. The presence of an industry data warehouse does not exclude operational transactional databases in the Data Layer. Most data will be manually entered into business and government systems. Gradually more data should be captured by sensor, with the data stored in transactional specialised databases and then transferred into the suggested industry data warehouse. The Persistence Layer manages the access points for data pulled from or pushed to other systems or platforms, and the data integration process. New web-portals and the generation of reports are managed at this layer. The collection and analysis tools from a commercially available system such as Crop Tracker, the pest management system (DAS) and coddling moth data portal (SIR) have to be in the Business Layer. The interfaces for the end-users, based on their roles and privacy settings, are managed in the top layer that delivers the data as information to the users.  The overall management of the industry information architecture should be the responsibility of the organization also responsible for the industry data warehouse.

The information architecture described here as layers of technology has the power to convert the industry capabilities from a CMMI Level 1 and 2 to CMMI Level 3 and beyond. It provides the industry with reliable information and knowledge for strategic decision-making, tracks operational data for market penetration and protection, and diffuses knowledge for continuous innovation.

# 4   Conclusions and recommendations

The results from this case study confirmed that the BC tree fruit industry is exposed to constantly changing environmental conditions that make effective production of tree fruits difficult.  Effective production in BC requires collective market positioning, whereby the chosen market determines some of the production processes across many growers and their packing-houses. This requires increasingly more complex data collection and processing across independent organizations. The most critical data domains for the BC tree fruit industry are with regards to 'pest and disease management' and 'fruit production, sales and marketing'.  Integration of data and knowledge management of these domains are necessary for success.  This study confirms that a technological solution for the BC tree fruit industry is available and recommends a process to transform an as yet unconnected industry cluster into one that can operate like a much larger integrated competitor would.

Research found in [21] showed that fragmented data and lack of connecting agricultural models with grower data is a common failing and limits progress.  This study now introduces a conceptual framework to start with assessing the industry's capability level for information management, confirms and prioritizes the missing digital data in the BC tree fruit industry.  CMMI levels of each of the data domains illustrate significant amounts of data, but no agreement of cross-industry processes and systems to guide data collection and data flows. Drilling down into the most significant gaps across stakeholder groups provides a roadmap for a technical solution to help the industry evolve.

Most agricultural innovation research offers a plethora of recommendations across multiple innovative tools, systems and techniques for precision agriculture.  This research focuses exclusively on information management of the BC tree fruit industry with data that is already available and suggests a process that prepares the industry for a future with sensors and sophisticated precision agriculture solution:

- Raw data capture efforts must be increased and integrated across the industry and supporting organizations.
- A separate industry data warehouse is needed with an extraction process from the farm management system(s) and other systems. The data has to be stored safely and securely, managed by a trusted industry stakeholder.
- Existing agricultural models and advisory services must be improved with real time data.
- Knowledge about existing and new markets must be acquired and integrated with production and data collection processes.

This can be achieved with a phased implementation plan whereby the industry initially only focuses on data capture processes and tools that encourage consistency in detail, frequency and quality of data, and the ability to integrate their data.  First, growers will have to agree and formalize the processes to capture data and collect it in a consistent format across the grower community. A key objective must be the integration of data in one architectural layer whereby the packed fruit can be traced back to its location on the orchard. The industry pest management system (DAS) must be connected to the orchard location and real time spay records. This then creates reliable historical data over time that can be made available at different levels of aggregation and anonymization for industry intermediaries and advisors.

In a later phase data management processes can evolve and use precision agriculture methods across the new integrated industry information architecture that now has the capability of knowledge creation, also enabling industry-wide controlled process improvements through collective decision-making. Based on good data, the industry decisions can include the roll out of new varieties and entry into new markets as a unified industry.  As seen in [10], there is an emphasis on the need to prove the value of big data. For an industry composed mainly of small and mid-size businesses, operating at varying levels of sophistication, the value of investing in big data is irrelevant in the absence of even the most basic data of reliable quality.  Building a data pool of integrated data and management processes for decision making and reflection is a prerequisite to any future industry ambitions and evaluation of options. The integration of systems and data sources, data analysis across the value chain and new predictive tools can place the BC Tree Fruit industry at a CMMI level 3 and beyond. With the time that it takes to collect historical data and implement change this is unlikely to happen before 2030.

Changes in agriculture and industry-wide transformations require long term planning. This case study seeks to provide an example of an industry that despite favourable conditions such as good local market access, supportive research institutions and available government funding experiences slow adoption of technology innovation and models. The findings illustrate that this is partly due to fundamental gaps in the industry information architecture which has not evolved in a robust and scalable manner. There is evidence that digitization and integration of systems by the grower community, whilst technically possible, is not happening. A process for assessment and mapping of an information architecture is described and a design roadmap is proposed. In the current environment of agricultural clusters that like the BC tree fruit industry lack growers and supporting organizations without the scale and capabilities for sophisticated information management solutions, similar assessments and mapping activities may have to be carried out. This can then assist in the creation of information management solutions for industry-wide collective decision making and co-innovation.

Findings, conclusions and recommendations from this case study cannot be generalised. However it is recommended that similar studies are carried out to assess the information capabilities of agricultural communities for co-innovation. This study focuses on one Western agricultural industry in the 21 century where it is easy to assume some degree of data management. Even if extensive data is collected, this may not be at the level that connects the industry stakeholders for competitive market positioning and adoption of precision agriculture. Connecting systems requires a data management foundation that allows integration and this has to be part of the architectural design for the industry from the outset. Much research effort has gone into the holistic approach to agricultural transformation and identifying the multitude of factors in agricultural innovation systems [2][27]. Although this is clearly beyond the scope of this paper, the social factors for industry change and decision-making will still have to be addressed and should not be underestimated for a complete analysis and a successful program of change for the BC tree fruit industry.

# 5 Acknowledgements

Funding for this research project was provided by the BC Innovation Fund and by the BC Government's Tree Fruit Competitiveness Fund; delivered by the BC Fruit Growers' Association and the Investment Agriculture Foundation of BC. The authors also wish to acknowledge the research support provided by staff of the British Columbia Fruit Growers' Association and student research assistants.